\documentclass[aps,pre,twocolumn,showpacs,preprintnumbers,amsmath,amssymb,superscriptaddress]{revtex4}
\usepackage{graphicx}
\usepackage{bm}
\usepackage{psfrag}
\usepackage{subfigure}

\newcommand{\be}{\begin{equation}}
\newcommand{\ee}{\end{equation}}
\newcommand{\ba}{\begin{eqnarray}}
\newcommand{\ea}{\end{eqnarray}}

\begin{document} 
\author{Matthieu Wyart}

\affiliation{Lewis-Sigler Institute, Princeton University, Princeton, NJ 08544;\\
Center for Soft Matter Research, New York University, NY 10003 }

\date{\today}

\title{Correlations between vibrational entropy and dynamics in liquids}

\begin{abstract}

A relation between vibrational entropy  and particles mean square displacement is derived in super-cooled liquids, assuming that the main effect of temperature changes is to rescale the vibrational spectrum. Deviations from this relation, in particular due to the presence of a Boson peak whose shape and frequency changes with temperature, are estimated. Using observations of the short-time dynamics in liquids  of various fragility, it is argued that (i) if the crystal entropy is significantly smaller than the liquid entropy at $T_g$, the extrapolation of the vibrational entropy leads to the correlation $T_K\approx T_0$, where $T_K$ is the Kauzmann temperature and $T_0$ is the temperature extracted from the Vogel-Fulcher fit of the viscosity. (ii) The jump in specific heat associated with vibrational entropy is very small for strong liquids, and increases with fragility. The analysis  suggests that these correlations  stem from the stiffening of the Boson peak under cooling, underlying the importance of this phenomenon on the dynamical arrest. 

\end{abstract}

\pacs{64.70.Pf,65.20.+w.77.22.-d}

\maketitle

When a liquid is cooled sufficiently rapidly to avoid crystallization, the relaxation time $\tau$ below which it behaves as a solid increases up to the glass transition temperature $T_g$ where the liquid falls out of equilibrium. In strong liquids $\tau$ displays an Arrhenius dependence on temperature, but in other liquids, said to be fragile, the  slowing-down of the dynamics is much more pronounced.  In general the dynamics is well captured by the Vogel-Fulcher law  $\log(\tau)=C+U/(T-T_0)$, although non-diverging functional forms can also reproduce the dynamics well \cite{mauro}. As the temperature evolves, two quantities appear to be good predictors of $\tau$:   the space available for the rattling of the particles on the picosecond time scales \cite{leporini,ngai},  embodied by the particles mean square displacement $\langle u^2\rangle$ observable in scattering experiments, and the difference between the liquid and the crystal entropy \cite{angellthermo}. When extrapolated below $T_g$ this quantity vanishes at some $T_K$,  the Kauzmann temperature \cite{kauzmann}, which appears to correspond rather well to the temperature $T_0$ extracted from the Vogel-Fulcher law \cite{tkt0}. The correlation between dynamics and thermodynamics has been interpreted early on as the signature of a thermodynamical transition at $T_K$ toward an ideal glass where the configurational entropy associated with the number of meta-stable states visited by the dynamics, or inherent structures, would vanish \cite{adam}. This view is appealing and still influential today \cite{wolynes,bouchaud}, although it is not devoid of conceptual problems \cite{stillinger}. Elastic models \cite{hall,dyre96,dyre} propose an alternative scenario of the glass transition: fragile liquids are simply those which stiffen under cooling, reducing the particle mean square displacement and increasing the activation barriers that must be overcome to flow.  In this view the rapid change of entropy in fragile liquids stem from the temperature dependence of the high-frequency elastic moduli \cite{granato}. This is consistent with some observations supporting that  fragile liquids stiffen more under cooling \cite{dyre96}, but predicts entropy variations in the liquid several times stronger than those observed \cite{dyre}.

Such distinct interpretations of the liquids entropy still coexist because the latter consists of several contributions  hard to disentangle experimentally \cite{goldstein}, in particular the configurational and the vibrational entropies. Phrased in the context of energy landscape, the vibrational entropy corresponds to the volume of phase space  associated with a typical configuration \cite{angellthermo}. The corresponding specific heat differs from the one of the phonon bath of a harmonic elastic network,  both due to the non-linearity of the interactions, and to the fact that the vibrational spectrum of the inherent structure evolves with temperature.  The latter effect  does not appear in high frequency dynamic heat capacities studies, as it requires changes of configurations that take place on the time scale $\tau$ to occur, but does contribute to the jump of specific heat characterizing the glass transition, whose amplitude is known to strongly correlate to the liquid fragility \cite{wang}. The configurational fraction  of the jump in specific heat has been estimated using calorimetry in quenched and annealed glasses \cite{goldstein}, measurements of elastic moduli \cite{Litovitz},  densities of states \cite{angellbp,Phillips,gjersing} and non-linear dielectric susceptibilities \cite{richert}. Overall, a great variation was found among glasses, with a configurational fraction ranging from 15$\%$  to $80\%$ and apparently decaying with the liquid fragility \cite{richert}.  Nevertheless, data are yet sparse and more analysis needs to be done to quantify respectively  the  correlations between dynamics and the different contributions to the entropy.  In this Letter  the relation between vibrational entropy and particle mean square displacement is investigated. From available data on the latter quantity, it is argued that the vibrational entropy by itself displays the evoked correlations with the dynamics, and that these correlations must be induced by the stiffening of some soft degrees of freedom under cooling, rather than via an overall rescaling of the vibrational spectrum.

On time scale $t<<\tau$ one may approximate a super-cooled liquid as a solid in a well-defined configuration. Scattering experiments yield a spectral analysis of the corresponding dynamics, enabling to define a density of states. The classical linear approximation for the vibrational entropy $S_{vib}$ per particle is then:
\be 
\label{1}
S_{vib}=\frac{1}{N} \sum_\omega k_B (1+\ln(\frac{k_B T}{\bar{h}\omega(T)}))
\ee
where $N$ is the number of particles, $\omega$ labels the frequency of the $3N$ modes, $T$ is the temperature, $k_B$ the Bolztmann constant. 
In Eq.(\ref{1}) the frequencies $\omega(T)$ are temperature dependent, both because configurations change with temperature, and because non-harmonicities are present within a configuration. The latter effect implies that Eq.(\ref{1}) is an approximation, the accuracy of which was tested using inelastic neutron scattering compared with calorimetric measurements in selenium \cite{Phillips}, where it was shown to be accurate throughout the glass phase. 
This leads for the vibrational  specific heat $C_{vib}$:
\be
\label{2}
C_{vib}= T\frac{\partial S_{vib}}{\partial T}=\frac{k_B}{N} \sum_\omega  (1-\frac{\partial \ln(\omega)}{\partial \ln(T)})= 3k_B (1-\langle \frac{\partial \ln(\omega)}{\partial \ln(T)}\rangle_\omega)
\ee
where $\langle X(\omega)\rangle_\omega\equiv \sum_\omega X(\omega)/(3N)$.
Several empirical studies have used scattering measurements to estimate the vibrational entropy via Eq.(\ref{1})  excluding non-linear effects by hyper-quenching \cite{angellbp} or not \cite{Phillips,gjersing}.
Unfortunately measuring the density of states at various temperature is difficult, although it is numerically feasible at limited viscosities \cite{mossa}, and  has not been done systematically for a broad range of fragility. 
On the other hand, it is intuitively clear that the vibrational entropy relates to the mean square displacement  of the particles $\langle u^2\rangle$ on short time scales \cite{dyre}, which indicates the space available for particles to fluctuate, and which is well studied experimentally. If the particles motions are de-correlated from each other, then the volume of phase space in this harmonic approximation is:
\be
\label{3}
\Omega=\prod_{i=1...N,\alpha} \langle \delta R_{i,\alpha}^2\rangle^{1/2}  \langle \delta P_{i,\alpha}^2\rangle^{1/2}/\bar{h} \propto \langle u^2\rangle^{3N/2} (m k_B T)^{3N/2} \notag
\ee
where $\alpha$ labels the three spatial coordinates, $\delta R_{i,\alpha}$ is the displacement of particle $i$ along the direction $\alpha$,  $\delta P_{i,\alpha}$ is the associated kinetic momentum and $m$ is the particle mass. The vibrational entropy is then:
\be
\label{4}
S_{vib}= k_B\ln(\Omega)=\frac{3}{2} k_B \ln(\langle u^2\rangle)+ S_0+ \frac{3}{2} k_B \ln(T)
\ee
where $S_0$ is a constant. This expression  in general over-estimates vibrational entropy, because particle motions are correlated:  $\langle \delta R_{i,\alpha} \delta R_{j,\beta} \rangle \neq 0$. Eq.(\ref{4}) is therefore an upper bound. 
This expression was discussed in \cite{dyre}, where it was shown to over-estimate at least by a factor two the total jump of specific heat of ortho-terphenyl. 
Eq.(\ref{4}) leads for the specific heat in this approximation:
\be
\label{4bis}
C_{vib}= \frac{3k_B}{2}+\frac{3k_BT}{2} \frac{\partial \ln(\langle u^2\rangle)}{\partial T}
\ee

The general relation can be obtained by considering the linear expression for the mean square displacement:
\be
\label{5}
\langle u^2\rangle\propto \frac{1}{N} \sum_\omega \frac{k_B T}{m \omega^2}
\ee
Differentiating Eq.(\ref{5}) with respect to temperature and using Eq.(\ref{2}) one gets:
\be
\label{6}
C_{vib} -\frac{3k_B}{2}=\alpha\frac{3k_BT}{2} \frac{\partial \ln(\langle u^2\rangle)}{\partial T}
\ee
where
\be
\label{7}
\alpha=\frac{1-2 \langle\frac{\partial \ln(\omega)}{\partial \ln(T)}\rangle_\omega}{1-2{\langle \frac{\partial \ln (\omega)}{\partial \ln(T)}\frac{1}{\omega^2}\rangle_\omega}/{\langle \frac{1}{\omega^2}\rangle_\omega}}
\ee

Thus Eq.(\ref{4bis}), is exact if the effect of temperature is to re-normalize all frequencies in the spectrum by the same factor, i.e. $\partial \ln(\omega)
/\partial T$ is independent of $\omega$, leading to $\alpha=1$ in Eq.(\ref{6}). 
Nevertheless there are obvious limitations in using Eq.(\ref{4bis}) as it stands to estimate the vibrational specific heat in molecular liquids.

(i) For molecules a fraction of the degrees of freedom are very stiff, and do not play a role near the glass transition (e.g. the vibrations of strong covalent bonds). 
Those degrees of freedom do not participate significantly to the mean square displacement of the particles. It has been proposed to describe molecules
as a set of independent beads \cite{lubchenko}, and to estimate the number of beads per molecule via the entropy of fusion, compared to the entropy of fusion in a hard sphere liquid $\tilde{s}_{HS}=1.16 k_B$.
This leads to a rough estimate for the number of beads per molecule $n_{bead}=(\Delta \tilde{H}/T_m)/ \tilde{s}_{HS}$, where $T_m$ and $\Delta \tilde{H}$ are respectively the fusion temperature and enthalpy.  Following this line of thought,  Eq.(\ref{4bis}) should be understood as the entropy per bead, rather than per atom. We shall use the notation $C_{vib}$ to designate specific heat per beads, and $\tilde{C}_{vib}$ for molar quantities. 

(ii) The assumption of overall shift of the typical interaction stiffness is in general violated. It is well known that, at least for fragile liquids, the shape of the spectrum changes with temperature. An excess of soft modes with respect to the Debye model for the density of states, the so-called Boson peak \cite{Phillipsbook}, shifts toward higher frequencies as the system is cooled \cite{tao,chumakov,angellbp}. Those modes presumably contribute significantly to the change of mean square displacement and to the vibrational specific heat \cite{angellbp,Phillips}.  If the increase of $\langle u^2\rangle$ under heating is mostly due to the softening of a limited number of low-frequency modes, the gain in entropy is diminished because the  short time scale dynamics becomes more correlated, as can be computed directly from Eq.(\ref{7}).  Using typical values for the Boson peak will lead to the estimate $1/5<\alpha<1$.  We shall in fact see below that $\alpha$ must be smaller than 1/2 for fragile liquids. Henceforth  I shall assume $\alpha$ to be constant and independent of the liquid fragility, and later come back on those assumptions and  estimate the effect of the Boson peak on $\alpha$.

To compute the vibrational entropy from Eq.(\ref{4bis}), one may use the observation that the $\alpha$-relaxation time scale $\tau$ and $\langle u^2\rangle$ appear to be universally related \cite{leporini}:
\be
\label{8}
\log(\tau)=f(\frac{\langle u_{Tg}^2\rangle}{\langle u^2\rangle})
\ee
where $\langle u_{Tg}^2\rangle$ is the mean square displacement at the glass transition $T_g$, extracted from the Debye-Waller factor measured at the vibrational (picosecond) time scale. In some liquids fast relaxation occurs at the nanosecond time scale \cite{alba}, which will presumably contribute to the total  entropy as well. This putative ``fast relaxational" entropy term is not incorporated in this analysis.  Note that  $\tau$ and $\langle u^2\rangle$ also appear to be strongly correlated in numerical studies of grain boundaries \cite{zhang}.  Eq.(\ref{8}) is observed to hold on a very wide range of liquids fragility, the scaling function found is $f(x)=\beta_0+\beta_1 x+\beta_2 x^2$ with $\beta_0=-0.424$, $\beta_1=1.622$ and $\beta_2=12.3$.  Using Eq.(\ref{6}) it is straightforward to obtain:
 \be
 \label{9}
 S_{vib}=S_1-\alpha\frac{3k_B}{2} \ln(\sqrt{\beta_1^2+4\beta_2(\log(\tau)-\beta_0)}-\beta_1)
 \ee
where $S_1$ is a constant. 
Replacing $\log(\tau)$ by its Volger-Fulcher expression $\log(\tau)=C+U/(T-T_0)$ in Eq.(\ref{9}) one obtains an expression for the vibrational entropy which diverges logarithmically at $T_0$ toward $-\infty$ (this is obviously unphysical and Eq.(\ref{8}) or the Volger-Fulcher must break down before $T_0$).  Under the assumption that the crystal entropy is sufficiently below the glass entropy near $T_g$, an extrapolation of the vibrational entropy and its logarithmic divergence must therefore cross the crystal entropy at a temperature $T_K$ near $T_0$. In this interpretation of the ($T_K$, $T_0$) correspondence, there is nothing specific about the crystal entropy itself, excepted that it is significantly lower than the entropy of the liquid. When this is not so, in particular in hard sphere liquids where the entropy of the liquid is lower than the entropy of the crystal above the melting pressure, $T_K=T_0$ (or the equivalent relation for packing fractions $\phi_K=\phi_0$) does not occur- $T_K$ is in fact ill-defined.  The relation between the apparently distinct temperatures $T_K$ and $T_0$ is generally interpreted as a strong support for the presence of a thermodynamic transition  where the configurational entropy would vanish\cite{wolynes}. Nevertheless  as soon as a relation between $\tau$ and $\langle u^2\rangle$ such as Eq.(\ref{8}) holds, those quantities are not independent, and an {\it apparent} divergence in $\tau$  must lead to an {\it apparent}  fall off in entropy at the same temperature, correlating $T_K$ and $T_0$- even thought no real divergences exist in this interpretation.

Below $T_g$, one may assume the system to remain in the same configuration. Neglecting  non-harmonic effects in the glass implies $\langle u^2\rangle\propto T$ leading to $C_{vib}= 3k_B$. This assumption appears to be a reasonable approximation for $T/T_g<0.8$ and up to $T_g$ in some glasses like polybutadiene, it nevertheless presumably overestimates the jump in specific heat in selenium where non-linearities are still significant just below $T_g$ \cite{ngai}. Together with Eq.(\ref{9}), this approximation leads to:
\be
\label{10}
\Delta C_{vib}=\frac{3k_B}{2}(\alpha \frac{m}{f'(1)}-1)
\ee
where $m=-\partial \log(\tau)/\partial\ln T|_{Tg}$ is the liquid fragility \cite{angellm} and $f'(1)\approx 26$. Eq.(\ref{10}) thus predicts the jump in vibrational specific heat to be strongly correlated to fragility. In particular for strong liquids with fragility in the twenties, even considering the maximal bound $\alpha=1$ one finds that $\Delta C_{vib}$ is essentially negligible. 
Experimentally it is found that fragility and molar jump of specific heat \cite{wang} satisfy (for non-polymeric glasses): $m=56 T_g \Delta\tilde{C}_p(T_g)/\Delta \tilde{H}$. From Eq.(\ref{10}) we get for the slope of the relation between fragility and jump in specific heat $m=52 \Delta  C_{vib}/(3\alpha k_B)= 52 \Delta \tilde{C}_{vib}/(3n_{bead} \alpha k_B)=\frac{20}{\alpha} T_m \Delta\tilde{C}_{vib}/\Delta \tilde{H}=\frac{30}{\alpha} T_g \Delta\tilde{C}_{vib}/\Delta \tilde{H}$, where the empirical rule $T_m=3/2 T_g$ was used. Thus if the jump in specific heat was fully vibrational one would have $\alpha=0.53\approx 1/2$. For fragile liquids  $\alpha$ cannot be significantly larger than that value, otherwise the jump of vibrational specific heat would be larger than the total  jump, which must also contain the positive contribution of the configurational entropy. 

I now come back to the value  and dependence of the parameter $\alpha$, which contains both the overall stiffening of the structure, and the softening of a limited number of degrees of freedom, the Boson peak. To estimate $\alpha$ in the case where only the Boson peak affects the mean square displacement, I consider the following approximation: the density of states is assumed to follow the Debye law $D(\omega)\propto \omega^2/\omega_D^3$ for $\omega<\omega_{BP}$, where $\omega_D$ is the Debye frequency. $D(\omega)$ remains of the same order $D(\omega)\propto 1/\omega_D$ for $\omega_{BP}<\omega<\omega_D$.  This corresponds to a Boson peak (maximum in $D(\omega)/\omega^2$) at $\omega_{BP}$. If $\omega_{BP}(T)$ is significantly smaller than $\omega_D$,  one finds  for such a spectrum $\langle 1/\omega^2\rangle_\omega\approx 1/(\omega_D \omega_{BP})$. Furthermore I assume that the modes near the Boson peak are those sensitive to the temperature, leading to $\langle \frac{\partial \ln (\omega)}{\partial \ln(T)}\frac{1}{\omega^2}\rangle_\omega\approx \langle \frac{\partial \ln (\omega)}{\partial \ln(T)}\rangle_\omega/\omega_{BP}^2$. Then Eq.(\ref{7}) gives: $\alpha=[1-2 \langle \frac{\partial \ln(\omega_{BP})}{\partial \ln(T)}\rangle_\omega]/[1-2\langle \frac{\partial \ln(\omega_{BP})}{\partial \ln(T)} \rangle_\omega \frac{\omega_D}{\omega_{BP}}]$.
In practice $d\ln(\omega_{BP})/d\ln(T) <0$ (the system softens with temperature), and one gets $\alpha>\omega_{BP}/\omega_{D}$. Typically values of the Boson peak in the glass are  $\omega_{BP}\approx \omega_{D}/5$ \cite{sokolov} leading to $\alpha>1/5$. From this primitive estimate of the lowest bound of $\alpha$, 
 one also gets that  $\alpha$ increases as the system is cooled, as does the Boson peak frequency, and is presumably smaller in fragile liquids where the shift of the Boson peak with temperature is pronounced. 
If $\alpha$ increases under cooling, the apparent divergence of the extrapolated vibrational entropy will only be enhanced, and the interpretation of the  $T_0=T_K$ correspondence 
remains safe.  On the other hand, the possible systematic decay of $\alpha$ with fragility may  diminish the predicted correlation between the vibrational part of the jump in specific heat and fragility.
Taking our lowest bound of $\alpha\approx 0.2 $ one still finds the vibrational contribution to the jump of specific heat to be significant (about 1/3) for the most fragile liquids. This  is consistent with the observations that the vibrational fraction of the jump of specific heat increases with fragility and represent more than half of the jump for fragile liquids \cite{richert}.


I have argued that as soon as a relation between relaxation time and mean square displacement  such as Eq.(\ref{8}) exists, dynamics and vibrational entropy must be correlated, leading to an alternative explanation for the $T_0=T_K$ correspondence. The amplitude of the correlations between fragility and jump in specific heat implies that the mechanism at play must include the softening under heating of a limited number of soft degrees of freedom, and is not simply a change of elastic moduli. This likely corresponds to the well-known softening of the Boson peak, which has been shown experimentally to affect significantly the thermodynamics of some liquids  \cite{angellbp}. Assuming that this is the case I find a reasonable estimate for the magnitude of the vibrational specific heat. 

It has been proposed since Adam-Gibbs \cite{adam,wolynes,bouchaud} that the relation between excess entropy and dynamics observed in liquids indicates that as less configurations are visited,  more collective rearrangements are required to relax the liquid, causing a rapid slowing down of the dynamics. However, there appears to be a viable alternative possibility consistent with this observation: when a liquid is cooled favored structures appear. This lowers the configurational entropy but more importantly stiffens the liquid which causes activation barriers to increase and vibrations to decrease. In this ``elastic" view of the glass transition the vibrational entropy most directly reflects the liquid dynamics, in consistence with the present analysis. Empirical evidences for such a stiffening range from vibrational spectra \cite{chumakov,tao,angellbp}, mean square displacements \cite{leporini,ngai} and elastic moduli \cite{exp,dyre96} measurements. In all cases strong correlations with the dynamics are found. Consistent with this approach, liquids at constant volume which are known to stiffen significantly less  \cite{angellbp,mossa} than at constant pressure are also much less fragile  \cite{angell2} in the cases known. Theoretically, the view that the stiffening of soft modes control the dynamics is supported among other works by the mode-coupling theory at moderate viscosities \cite{gotze,parisi}, by geometric arguments on the nature of the Boson peak and pre-vitrification in hard sphere liquids \cite{brito} and by the numerical observations that activated events required to flow occur mostly along soft modes belonging to the foot of the Peak \cite{brito2,harrowell}.  Theoretical efforts required to assess this view further should focus in particular on the still obscure relation between excess soft modes and viscosity in highly viscous liquids, and on the questions of what in its microscopic structure  determine if a liquid stiffens or not under cooling.

It is a pleasure to thank C.A. Angell, M. Cates, G. Tarjus  and F. Zamponi for comments on the manuscript.


\begin{thebibliography}{99}

\bibitem{mauro} J.C. Mauro, Y. Yue, A.J. Ellison, P.K. Gupta \& D.C. Allan, PNAS {\bf 106}, 19780 (2009); N.Hecksher, A.I. Nielsen, N. B. Olsen \& J.C. Dyre, Nat. Phys. {\bf 4} 737 (2008)

\bibitem{leporini} L. Larini, A. Ottochian, C De Michele and D. Leporini, Nature Phys., {\bf 4}, 42 (2008).

\bibitem{ngai} K.L. Ngai, Philo. Mag., {\bf 84},1341 (2004)

\bibitem{angellthermo} L-M. Martinez and C.A. Angell, Nature, {\bf 410}, 663 (2001).

\bibitem{kauzmann} W. Kauzmann, Chem. Rev. {\bf 43}, 219 (1948).

\bibitem{tkt0} C.A. Angell, J. Res. Natl. Inst. St. Tech., {\bf 102}, 171 (1997)

\bibitem{adam} A. Adam and J.H. Gibbs, J. Chem. Phys., {\bf 43}, 139 (1965).

\bibitem{wolynes} V. Lubchenko and P. Wolynes, Annu. Rev. Phys. Chem., {\bf 58}, 235-266 (2007).

\bibitem{bouchaud} J-P Bouchaud and G. Biroli. J. Ch. Phy. {\bf 121}, 7347 (2004)

\bibitem{stillinger} F. Stillinger, J. Chem. Phys. {\bf 88}, 7818, (1988)

\bibitem{hall} R.W. Hall and P. Wolynes, J. Ch. Phys. {\bf 86}, 2943 (1987)

\bibitem{dyre96} J.C. Dyre, N.B. Olsen and T. Christensen, Phys. Rev. B, {\bf 53}, 2171 (1996).

\bibitem{dyre} J.C. Dyre, Rev. of Mod. Phys., {\bf 78}, 3, (2006).

\bibitem{granato} A. V. Granato, J. of Non-Cry. Sol.,{\bf 307-310}, 376 (2002).


\bibitem{goldstein} M. Goldstein, J. Chem. Phys. {\bf 64}, 11, (1976).

\bibitem{wang} L-M Wang, C.A. Angell and R. Richert, J. Chem. Phys. {\bf 125}, 074505, (2006).

\bibitem{Litovitz} TA. Litovitz,  Jour. Acoust. Soc. Am. {\bf 31}, 681, (1959).

\bibitem{angellbp} C.A. Angell, Y. Z. Yue,  L-M Wang, J.R.D. Copley, S. Borick and S. Mossa, J. Phys. Cond. Mat. {\bf 15} S1051 (2003);A.I.P Conf. Pro., slow dyn. 2003, {\bf 708}, 473 (2004)

\bibitem{Phillips} W.A. Phillips, U. Buchenau, N. Nucker, A-J Dianoux and W. Petry, Phys. Rev. Lett., {\bf 63}, 2381, (1989).

\bibitem{gjersing} E.L. Gjersing, S. Sen, B.G. Aitken, Jour. Non-Cryst. Solids, {\bf 355}, 748 (2009).

\bibitem{richert} L-M Wang and R. Richert, Phys. Rev. Lett, {\bf 99}, 185701, (2007).

\bibitem{Phillipsbook} {\it Amorphous Solids. Low Temperature Properties}, edited by  W. A.  Phillips, Springer-Verlag, Berlin (1981).   

\bibitem{mossa} S. Mossa et al., Phys. Rev. E, {\bf 65}, 041205

\bibitem{chumakov} A.I. Chumakov, I. Sergueev, U. van Burck, W. Schirmacher, T. Asthalter, R.Ruffer, O. Leupold and W. Petry, Phys. Rev. Lett., {\bf 92}, 245508 (2004).

\bibitem{tao} N.J. Tao, G. Li, X. Chen, W.M. Du and H.Z. Cummins, Phys. Rev. A, {\bf 44}, 6665 (1991).

\bibitem{lubchenko} V. Lubchenko and P. Wolynes, J. Chem. Phys. {\bf 119}, 17, (2003).

\bibitem{angellm} C.A. Angell, Science, {\bf 267}, 1924 (1995).

\bibitem{zhang} H. Zhang, D. Srolovitz, J.F. Douglas and J.A. Warren, PNAS, {\bf 106} 7735 (2009)

\bibitem{alba} K. Niss, C. Dalle-Ferrier, B. Frick, D. Russo, J. Dyre and C. Alba-Simionesco arXiv:0908.2046 (2009)


\bibitem{sokolov} V. K. Malinovsky et al., J. Phys. Cond. Mat. {\bf 4} 139 (1992)

\bibitem{exp} D. H. Torchinsky, J. A. Johnson, and K. A. Nelson, Jour. of Chem. Phys., {\bf 130}, 064502 (2009)



\bibitem{angell2} C.A. Angell , S. Borick, Jour. of Non-Cryst. Sol. {\bf 307Ð310} 393Ð40 (2002) 

\bibitem{gotze} W. Gotze and M.R. Mayr, Phys. Rev. E {\bf 61}, 587 (2000).

\bibitem{parisi} G. Parisi,  Eur. Phys. Jour.  E, {\bf 9},  213-218 (2002). 

\bibitem{brito} C. Brito and M. Wyart, J. Chem. Phys. {\bf 131}, 024504 (2009)

\bibitem{brito2} C. Brito and M. Wyart, J. Stat. Mech.: Theory Exp. 2007, L08003.

\bibitem{harrowell} A. Widmer-Cooper, H. Pierry, P. Harrowell, and D. Reichman, Nat. Phys.  {\bf 4}, 711, 2008.













\end{thebibliography}
\end{document}